\documentclass[11pt,a4paper]{article}

\usepackage{amsmath}
\usepackage{amsthm,amssymb}

\usepackage{graphicx} 

\newcommand{\beq}{\begin{equation}}
\newcommand{\eeq}{\end{equation}}

\numberwithin{equation}{section}

\begin{document}
\title{Exact solution of a model of a vesicle attached to a wall subject to mechanical deformation}
\author{A L Owczarek$^1$ and T Prellberg$^2$\\
  \footnotesize
  \begin{minipage}{13cm}
    $^1$ Department of Mathematics and Statistics,\\
    The University of Melbourne, Parkville, Vic 3010, Australia.\\
    \texttt{owczarek@unimelb.edu.au}\\[1ex] 
$^2$ School of Mathematical Sciences\\
Queen Mary University of London\\
Mile End Road, London E1 4NS, UK\\
\texttt{t.prellberg@qmul.ac.uk}
\end{minipage}
}

\maketitle  

\begin{abstract}

Area-weighted Dyck-paths are a two-dimensional model for vesicles attached to a wall.
We model the mechanical response of a vesicle to a pulling force by extending this model.

We obtain an exact solution using two different approaches, leading to a $q$-deformation
of an algebraic functional equation, and a $q$-deformation of a linear functional equation
with a catalytic variable, respectively. 
While the non-deformed linear functional equation is solved by substitution of special values
of the catalytic variable (the so-called ``kernel method''), the $q$-deformed case is solved
by iterative substitution of the catalytic variable.

Our model shows a non-trivial phase transition when a pulling force is applied. As soon as the
area is weighted with non-unity weight, this transition vanishes.

\end{abstract}

\section{Introduction}
With the growing experimental ability to probe the behaviour of a single polymer under the influence
of a stretching force \cite{haupt1999,hauchke2004}, there has been an effort by theoretical work to 
explore models that cover the possible scenarios encountered. For example, various directed path models of a polymer with one end, or both ends, grafted to a surface and being pulled at the end or in the middle, with the addition of an attractive contact interaction with the surface, have been studied in two and three 
dimensions \cite{orlandini2004a,orlandini2004b,brak2009,alvarez2009,rensburg2010}. Numerical work on 
self-avoiding walks models has also confirmed and expanded upon the predictions of these directed 
models \cite{krawczyk2004,krawczyk2005,mishra2005,orlandini2009}. In the continuum limit, this problem is related to the area under a Brownian curve \cite{majumdar2005}.

Another type of model, that of a lattice polygon that has been weighted by its length and area, 
can rather be used to model biological vesicles 
\cite{leibler1987,brak1990a-a,fisher1991a-a,owczarek1993a-:a,brak1994a-:a,brak1998}. For example, the effect of a restricted geometry such as a slit has recently been analysed \cite{owczarek2010a-:a}.
Here, it is also natural to consider the effect of a pulling force of the surface of the vesicle.

Pulling membrane-anchored microspheres using optical traps is a technique commonly used in experimental settings \cite{settles2010}. When preparing the sample, microspheres can bind anywhere on the membrane, and due to small anisotropies in the experimental setup perfect vertical pulling is difficult to achieve. 

Hence, we study a simple directed model of a two-dimensional vesicle that is being pulled away from a wall. The 
anisotropy is modelled by a pulling force with a small horizontal component, which leads to the force acting on the
right-most peak of the vesicle. Only the vertical component of the force contributes to the configurational energy of the vesicle.

The model is of interest physically as well as mathematically.  It illustrates the
solution of functional equations with catalytic variables under a $q$-deformation, when the solution gets expressed in terms of $q$-series rather than in terms of algebraic functions.

\section{The Model}

\begin{figure}[ht!]
\begin{center}\includegraphics[width=0.8\textwidth]{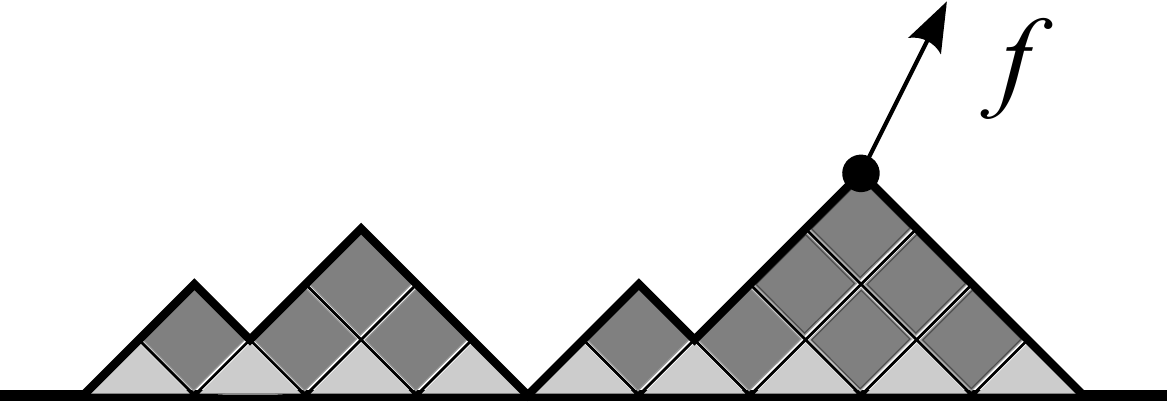}\end{center}
\caption{An area-weighted Dyck path with $18$ steps and area of size $11$, which is being pulled away 
from the surface at its right-most peak with force $f$. The height $h$ of the right-most peak in this 
example is $4$.
\label{figure1}
}
\end{figure}

Dyck paths are directed walks on $\mathbb{Z}^2$ starting at $(0,0)$ 
and ending on the line $y=0$, which have no vertices with negative
$y$-coordinates, and which have steps in the $(1,1)$ (up-step) and $(1,-1)$ (down-step)
directions \cite{stanley1999}. 
Given a Dyck path $\pi$, we define the length $n(\pi)$ to be half the number of
its steps, and the area $m(\pi)$ to be the sum of the starting heights of all up-steps
in the path. This is sometimes also called the rank function
of a Dyck path \cite{ferrari2005,sapounakis2006}, and is equivalent to the number of diamond
plaquettes under the Dyck path. We further define the height $h(\pi)$ of the right-most peak
as the length of the final descent of the Dyck path, i.e. the maximal number of the last
consecutive down-steps. An example of a Dyck path is given in Fig.~\ref{figure1}.

Let ${\mathcal{G}}$ be the set of Dyck paths, and define
the generating function
\beq
G(t,q,\lambda)=\sum_{\pi\in\mathcal{G}}t^{n(\pi)}q^{m(\pi)}\lambda^{h(\pi)}\;. 
\eeq
The generating function for area-weighted Dyck paths \cite{flajolet1980,duchon2000} has previously been studied.
Dyck paths weighted by height of the right-most peak are related to Ballot paths,
which have been previously studied as a model of pulled polymers \cite{orlandini2004a}. Here we extend 
these works by combining and extending these models.

The key to obtaining a solution of this model is a combinatorial construction of Dyck
paths which leads to a functional equation for the generating function $G$. We shall present
two different constructions leading to two structurally different functional equations.

\section{$q$-deformed Algebraic Functional Equation}

\begin{figure}[ht!]
\begin{center}\includegraphics[width=0.6\textwidth]{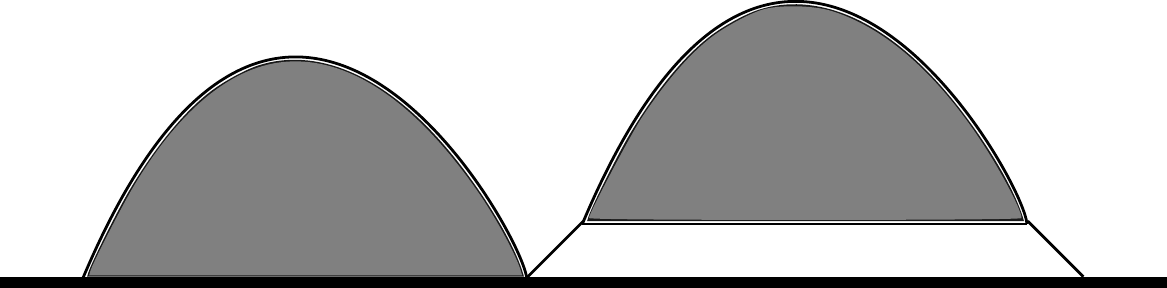}\end{center}
\caption{A schematic decomposition leading to Eqn. (\ref{q-algebraic-lambda}): A non-zero Dyck path decomposes uniquely into a Dyck path, followed by another Dyck path which is bracketed by up and down steps.
\label{decomposition1}
}
\end{figure}

To obtain the first functional equation, we decompose Dyck paths by considering the right-most up-step 
starting at height zero. This leads to a decomposition of a Dyck path in terms of two smaller
Dyck paths.  More precisely, except for the zero-step path, this decomposes a Dyck path uniquely 
into a Dyck path, followed by another Dyck path which is bracketed by up and down steps, as indicated
in Figure \ref{decomposition1}.

This naturally leads to a functional equation which is quadratic in the generating functions,
\beq
\label{q-algebraic-lambda}
G(t,q,\lambda)=1+\lambda tG(t,q,1)G(qt,q,\lambda)\;.
\eeq
Note that in the product $\lambda tG(t,q,1)G(qt,q,\lambda)$ the left Dyck path no longer contributes to the
height of the right-most peak, leading to the factor $G(t,q,1)$, and the bracketing of the right 
Dyck path by up and down steps increases the area proportional to the length of the path, as well as increases height and
length by one, leading to the factor $\lambda tG(qt,q,\lambda)$.

\subsection{The Case $q=1$}

The functional equation (\ref{q-algebraic-lambda}) reads now
\beq
\label{algebraic-lambda}
G(t,1,\lambda)=1+\lambda tG(t,1,1)G(t,1,\lambda)\;,
\eeq
which after substitution of $\lambda=1$ can be readily solved to give
\beq
\label{Gt11}
G(t,1,1)=\frac{1-\sqrt{1-4t}}{2t}=\frac2{1+\sqrt{1-4t}}\;,
\eeq
which is the well-known generating function for Catalan numbers.
Substitution of this result back into Eqn.~(\ref{algebraic-lambda}) gives
\beq
\label{Gt1l}
G(t,1,\lambda)=\frac2{2-\lambda+\lambda\sqrt{1-4t}}\;.
\eeq
For comparison with the result in Eqn.~(\ref{first-solution}), note that
a series expansion in $\lambda$ gives
\beq
\label{above}
G(t,1,\lambda)=\sum_{l=0}^\infty\lambda^lt^lG(t,1,1)^l\;. 
\eeq

\subsection{The case $q\neq1$}

Substituting $\lambda=1$ into the functional equation (\ref{q-algebraic-lambda}) gives
\beq
\label{q-algebraic}
G(t,q,1)=1+tG(t,q,1)G(qt,q,1)\;.
\eeq
The Ansatz
\beq
G(t,q,1)=\frac{H(qt,q)}{H(t,q)}
\eeq
gives a linear functional equation for $H(t,q)$,
\beq
\label{linearfeqn}
H(qt,q)=H(t,q)+tH(q^2t,q)\;.
\eeq
Inserting $H(t,q)=\sum_{n=0}^\infty c_n(q)t^n$ into (\ref{linearfeqn}) gives a recurrence
$q^nc_n(q)=c_n(q)+q^{2(n-1)}c_{n-1}(q)$ for the coefficients, which can be readily solved by iteration. Specialising to
$c_0(q)=1$, we find
\beq 
\label{solnlinearfeqn}
H(t,q)=\sum_{k=0}^\infty\frac{(-t)^kq^{k(k-1)}}{(q;q)_k}\;,
\eeq
where we have used the $q$-product notation
\beq
(t;q)_k=\prod_{j=0}^{k-1}(1-tq^j)\;.
\eeq
Here $H(t,q)={}_0\phi_1(-;0;q,-t)$ is a basic hypergeometric series \cite{gasper2004} also referred to as the $q$-Airy function
\cite{ismail2005}.

Once $G(t,q,1)$ is known and substituted back into Eqn.~(\ref{q-algebraic-lambda}), we can
solve for $G(t,q,\lambda)$ by iteration and obtain
\beq
\label{first-solution}
G(t,q,\lambda)=\sum_{l=0}^\infty\lambda^lt^lq^{\binom l2}\frac{H(q^lt,q)}{H(t,q)}\;.
\eeq
This iteration makes sense in a ring of formal power series. Alternatively, one can argue that
the iteration converges for $|q|<1$.

\begin{figure}[ht!]
\begin{center}\includegraphics[width=0.8\textwidth]{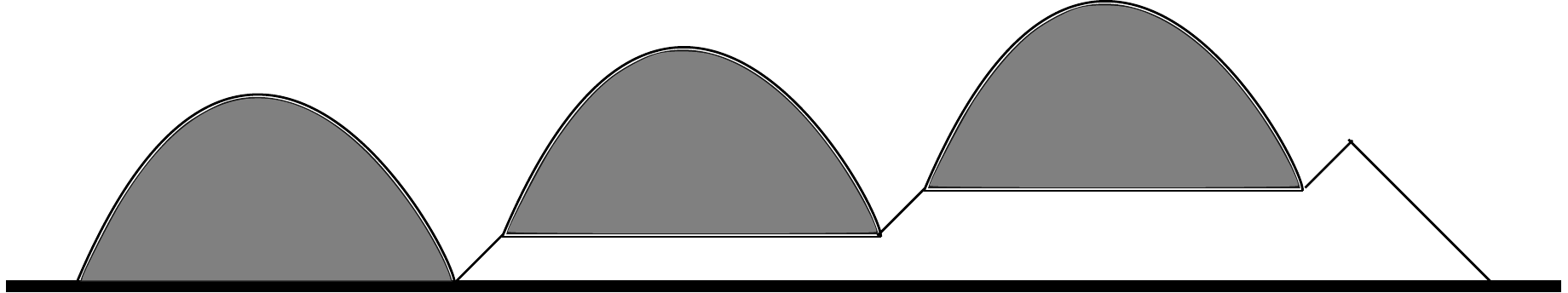}\end{center}
\caption{The term $l=3$ in the sum of (\ref{first-solution}) corresponds to the concatenation
of three Dyck paths separated by single up-steps, followed by a hook formed of one up-step and three down-steps.
\label{combinatorics}
}
\end{figure}

We note that this solution also admits a direct combinatorial interpretation. The $l$-th term in the sum
of (\ref{first-solution}) corresponds to the concatenation of $l$ Dyck paths separated by single
up-steps, followed by a final hook formed by one up-step and $l$ down-steps. The generating function for
a Dyck path starting at height $l$ is given by $G(q^lt,q,1)$, and the final descent of length $l$ increases the 
area by $\binom l2$, leading to an overall weight of
\beq
G(t,q,1)tG(qt,q,1)t\cdot\ldots\cdot G(q^lt,q,1)t\lambda^lq^{\binom l2}=\lambda^lt^lq^{\binom l2}\frac{H(q^lt,q)}{H(t,q)}\;.
\eeq
Figure \ref{combinatorics} indicates this for the case $l=3$.

Finally, we note that the limit $q\to1$ can be obtained as follows. As 
\beq
G(t,1,1)=\lim_{q\to1}G(t,q,1)=\lim_{q\to1}\frac{H(qt,q)}{H(t,q)}\;, 
\eeq
we deduce that for fixed value of $k$, 
\beq
G(t,1,1)=\lim_{q\to1}\frac{H(q^{k+1}t,q)}{H(q^kt,q)}
\eeq
and thus
\beq
\lim_{q\to1}\frac{H(q^lt,q)}{H(q,t)}=G(t,1,1)^l\;.
\eeq
Therefore Eqn.~(\ref{first-solution}) reduces to the result in Eqn.~(\ref{above}) as $q\to1$.

\section{$q$-deformed Linear Functional Equation}

To obtain the second functional equation, we consider how Dyck path configurations get changed
by the insertion of a small pattern. 

The right-most descent, together with the preceding up-step forms a \emph{hook}. Inserting an
up-step followed immediately by a down-step in the right-most descent creates a new Dyck path 
with a new hook. If the descent has length $h$, there are $h+1$ possibilities of insertion.

\begin{figure}[ht!]
\begin{center}\includegraphics[width=0.8\textwidth]{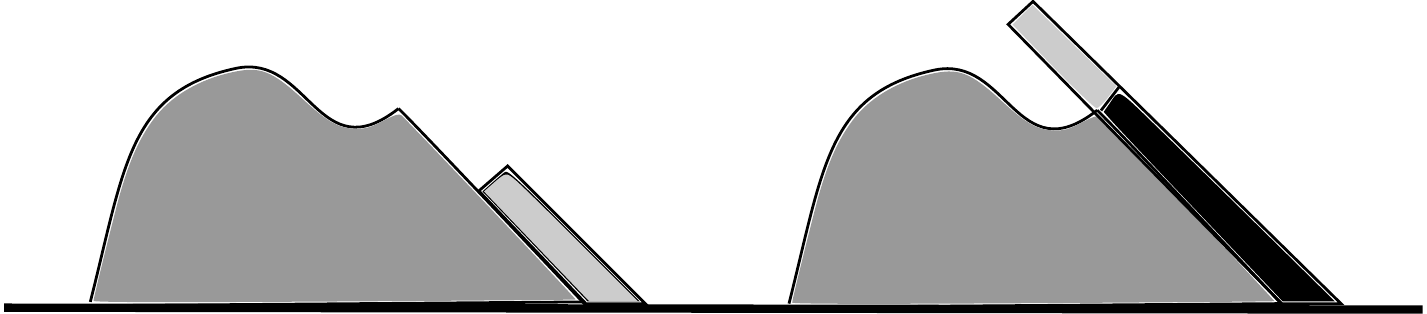}\end{center}
\caption{A schematic construction leading to Eqn. (\ref{q-linear-lambda}): Inserting an
up-step followed immediately by a down-step in the right-most descent creates a new Dyck path 
with a new hook (left). When adding hooks of all possible lengths, one needs to correct for
overhangs produced by hooks that are too long (right).
\label{decomposition2}
}
\end{figure}

The consequence of this insertion is the addition of a column to the Dyck path as indicated on
the left-hand-side of Figure \ref{decomposition2}.

The functional equation corresponding to this construction is given by
\beq
\label{q-linear-lambda}
G(t,q,\lambda)=1+\frac{\lambda t}{1-q\lambda}G(t,q,1)-\frac{q\lambda^2t}{1-q\lambda}G(t,q,q\lambda)\;.
\eeq

The term $\frac{\lambda t}{1-q\lambda}G(t,q,1)$ corresponds to the addition of a column of arbitrary height to the Dyck path. The ensuing overcounting is compensated for by subtraction of the term $\frac{q\lambda^2t}{1-q\lambda}G(t,q,q\lambda)$, which corresponds to the removal of the columns that produce invalid configurations, as indicated on the right-hand-side of Figure \ref{decomposition2}.

Alternatively, the terms on the right-hand side of (\ref{q-linear-lambda}) can be understood by considering 
\beq
G(t,q,\lambda)=\sum_{h=0}^\infty\lambda^hG_h(t,q)\;,
\eeq
where $G_h(t,q)$ are partial generating functions for fixed height $h$. The only Dyck path with a hook of height $h=0$ is
a zero-step path. A Dyck path with a hook of height $h\geq1$ can only have been constructed from a Dyck path with a hook of 
height at least $h-1$. By the construction, its area increases by an amount $h-1$, and we thus find that for $h\geq1$
\beq
G_h(t,q)=q^{h-1}\sum_{k=h-1}^\infty G_k(t,q)\;.
\eeq
Summing over the partial generating functions leads to the functional equation (\ref{q-linear-lambda}).

\subsection{The Case $q=1$}

The functional equation (\ref{q-linear-lambda}) now reads
\beq
\label{linear-lambda}
G(t,1,\lambda)=1+\frac{\lambda t}{1-\lambda}G(t,1,1)-\frac{\lambda^2t}{1-\lambda}G(t,1,\lambda)\;.
\eeq
Here, substitution of $\lambda=1$ is no longer useful, and one solves the functional equation using the
kernel method \cite{prodinger2004}. Writing Eqn.~(\ref{q-linear-lambda}) as
\beq
\label{kernel-equation}
K(t,\lambda)G(t,1,\lambda)=1+\frac{\lambda t}{1-\lambda}G(t,1,1)\;,
\eeq
where the kernel $K(t,\lambda)$ is given as
\beq
\label{kernel}
K(t,\lambda)=1+\frac{\lambda^2t}{1-\lambda}\;,
\eeq
we observe that the left-hand side of Eqn.~(\ref{kernel-equation}) vanishes if $\lambda=\lambda_0$ is
chosen such that $K(t,\lambda_0)$ vanishes (provided $G(t,1,\lambda_0)$ does not diverge). This then implies
that
\beq
G(t,1,1)=\frac{\lambda_0-1}{\lambda_0t}\;,
\eeq
and substituting this result back into Eqn.~(\ref{kernel-equation}) leads to an expression for $G(t,1,\lambda)$.
The appropriate root of the kernel is given by
\beq
\lambda_0=\frac{1-\sqrt{1-4t}}{2t}\;,
\eeq
Using this value, we recover the expressions (\ref{Gt11}) for $G(t,1,1)$ and (\ref{Gt1l}) for $G(t,1,\lambda)$ given
above.

\subsection{The Case $q\neq1$}

If $q\neq1$, we can solve the functional equation (\ref{q-linear-lambda}) by iterative substitution. Substituting
$\lambda q^k$ for $\lambda$ in (\ref{q-linear-lambda}), we find
\beq
\label{q^k-linear-lambda}
G(t,q,q^k\lambda)=1+\frac{q^k\lambda t}{1-q^{k+1}\lambda}G(t,q,1)-\frac{q^{2k+1}\lambda^2t}{1-q^{k+1}\lambda}G(t,q,q^{k+1}\lambda)\;.
\eeq
We can therefore express $G(t,q,q^k\lambda)$ in terms of $G(t,q,q^{k+1}\lambda)$ and use this to iterate. As
a result, we find
\beq
G(t,q,\lambda)=\sum_{k=0}^\infty\frac{(-\lambda^2t)^kq^{k^2}}{(q\lambda;q)_k}\left(1+\frac{q^k\lambda t}{1-q^{k+1}\lambda}G(t,q,1)\right)\;.
\eeq
Defining
\beq
I(t,q,y)=\sum_{k=0}^\infty\frac{(-t)^kq^{k(k-1)}}{(y;q)_k}\;,
\eeq
this simplifies to
\beq
\label{nearlythere}
G(t,q,\lambda)=I(q\lambda^2t,q,q\lambda)-\frac1\lambda\left[I(\lambda^2t,q,q\lambda)-1\right]G(t,q,1)\;.
\eeq
Note that $I(t,q,y)={}_1\phi_2(q;0,y;q,-t)$ is also expressible as a basic hypergeometric function. Moreover,
it a deformation of $H(t,q)$, in that $I(t,q,q)=H(t,q)$.

Substituting $\lambda=1$ in Eqn.~(\ref{nearlythere}) gives now immediately
\beq
G(t,q,1)=\frac{H(qt,q)}{H(t,q)}\;,
\eeq
a result already obtained from the $q$-algebraic functional equation (\ref{q-algebraic}) above. Substituting
this result back into Eqn.~(\ref{nearlythere}) leads to the main result of this section, namely
\beq
\label{second-solution}
G(t,q,\lambda)=I(q\lambda^2t,q,q\lambda)-\frac1\lambda\left[I(\lambda^2t,q,q\lambda)-1\right]\frac{H(qt,q)}{H(t,q)}\;.
\eeq
Note the rather different structure of the equivalent solutions (\ref{first-solution}) and (\ref{second-solution}). 
The former is in terms of an infinite sum, which however readily allows for extraction of the partial generating 
function for fixed height $h$. The latter is more compact as it is written in terms of a finite number of special 
functions. We note that the equivalence of both solutions implies the identity
\beq
\sum_{l=0}^\infty\lambda^lt^lq^{\binom l2}H(q^lt,q)=
I(q\lambda^2t,q,q\lambda)H(t,q)-\frac1\lambda\left[I(\lambda^2t,q,q\lambda)-1\right]H(qt,q)\;.
\eeq

\section{Analysis of behaviour}

\begin{figure}[ht!]
\begin{center}\includegraphics[width=0.6\textwidth]{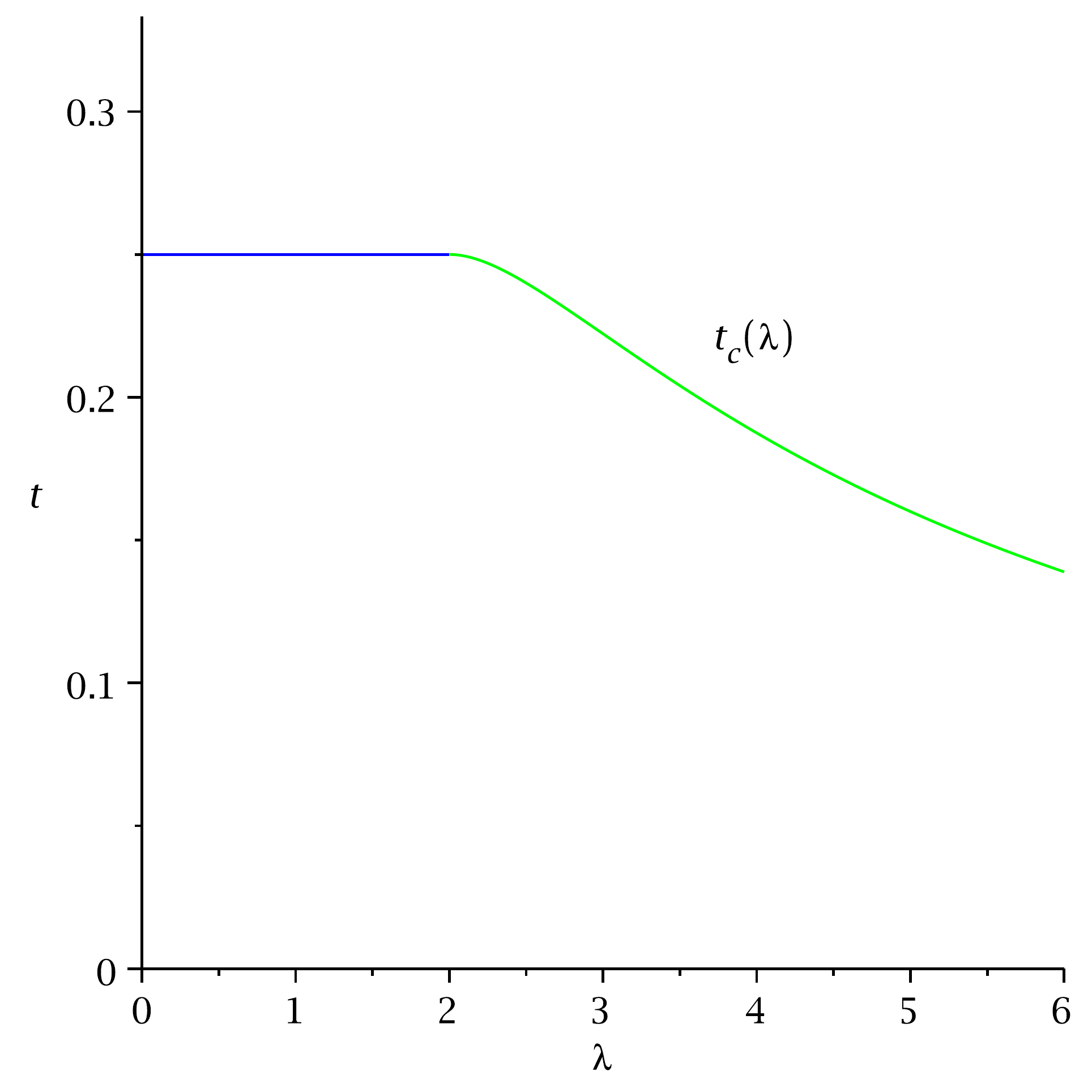}\end{center}
\caption{A plot of the radius of convergence $t_c(\lambda)\equiv t_c(\lambda,1)$ in the polymer length variable $t$ 
of the generating function when $q=1$ as a function of the force variable $\lambda$. Note that there is a change of
behaviour at $\lambda=2$ with constant $t_c(\lambda)=1/4$ for $\lambda\leq2$.
\label{figure2}
}
\end{figure}

Let us begin by considering the case of $q=1$, that is no extra area weighting or, equivalently, osmotic 
pressure. The first important point to make is that previous models of pulling on the end of a polymer 
without any other interaction with a surface give the result that as soon as a force is applied the end 
of the polymer changes its scaling behaviour from random-walk scaling
\begin{equation}
\langle h\rangle \sim n^{1/2}
\end{equation}
for $f=0$ ($\lambda=1$) to ballistic scaling
\begin{equation}
\langle h\rangle \sim n
\end{equation}
for $f>0$ ($\lambda>1$). 
This is reflected in the fact that the radius of convergence of the associated generating function is constant 
for $\lambda\leq1$ and decreases continuously for $\lambda>1$. 
As a model of pulling on a vesicle attached to a surface, our model has the feature that the pulling occurs on the 
final ``peak'', the location of which along the surface can change with the pulling itself. 
As such, there is a non-trivial critical point as a function of $\lambda$ in our model, with a critical value
\begin{equation}
\lambda = \lambda_c \equiv 2\;.
\end{equation}
A plot of the radius of convergence of the generating function~(\ref{Gt1l}) is shown in Fig.~\ref{figure2}.

Note that the reduced free energy $\kappa(\lambda,q)$ of the surface is simply related to the radius of 
convergence $t_c(\lambda,q)$ in the usual way via
\begin{equation}
\kappa(\lambda,q) = - \log t_c(\lambda,q)\;.
\end{equation}

For $q<1$ the generating function~(\ref{first-solution}) can be seen to be convergent except 
where $H(t,q)=0$. Crucially this is independent of $\lambda$! This implies that for $q<1$ there 
is no phase transition as a function of $\lambda$, since $\kappa(\lambda,q)$ is independent of 
$\lambda$ for $q<1$. Physically this means that the osmotic pressure always dominates the pulling 
force for $q<1$: $\langle h\rangle$ will be bounded for all $\lambda$.

\begin{figure}[ht!]
\begin{center}\includegraphics[width=0.6\textwidth]{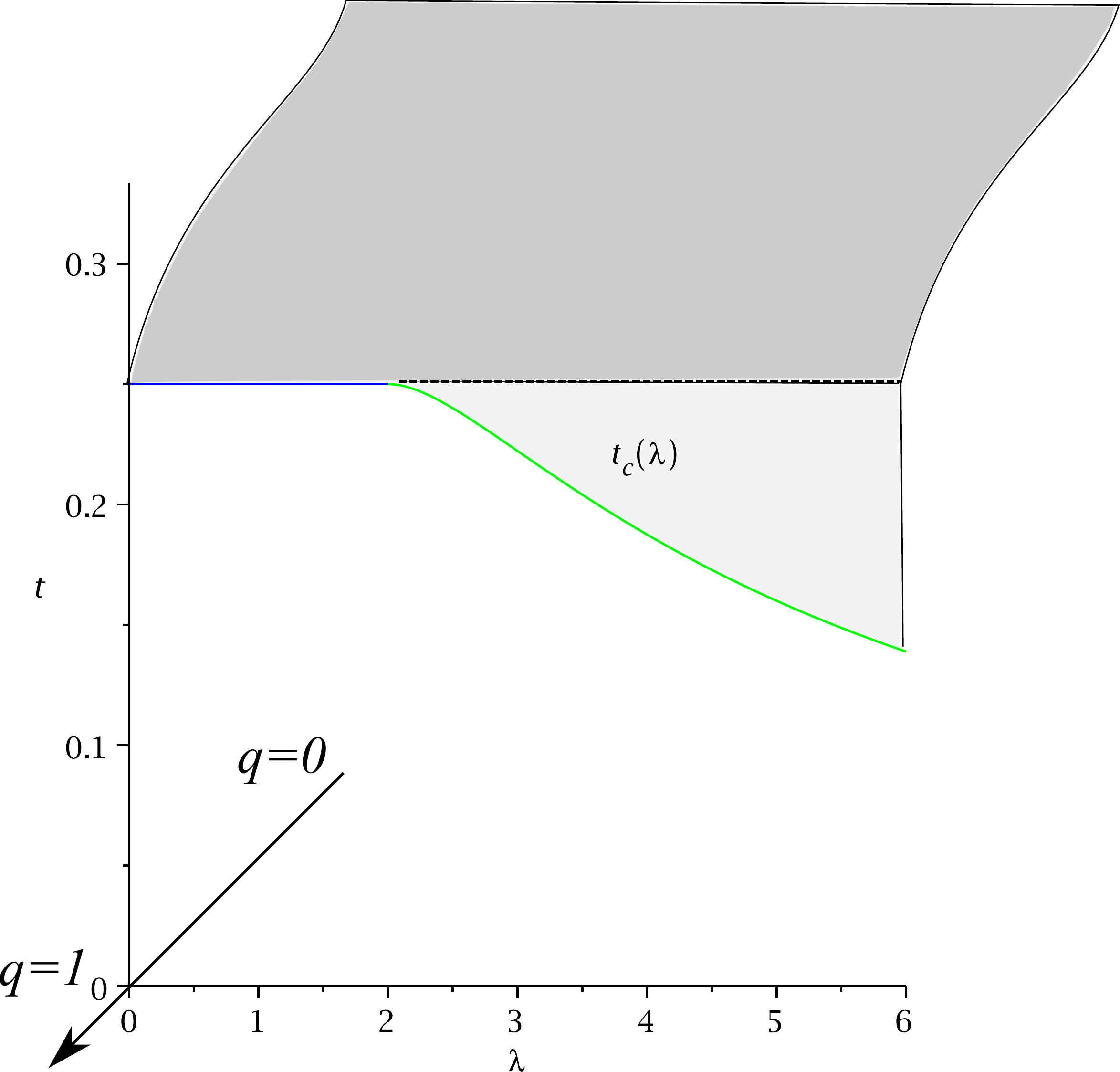}\end{center}
\caption{A plot of the radius of convergence $t_c(\lambda,q)$ in the polymer length variable $t$ 
of the generating function for general $q$ as a function of the force variable $\lambda$. Note that $t_c(\lambda,q)$ is independent of $q$ for $q<1$ and hence that there is a jump of $t_c(\lambda,q)$ as $q$ approaches one from below.
\label{figure3}
}
\end{figure}

Interestingly, if one considers the limit $q\rightarrow 1$ for $\lambda>2$ one is left with the problem 
that the free energy ``jumps'' at $q=1$ as indicated in Figure \ref{figure3}. The resolution of this apparent problem is that for $q>1$ one 
needs to consider the free energy per unit area rather than the free energy per unit length of the surface. 
The limiting free energy per unit area is a continuous function of $q$ and $\lambda$ for all $\lambda$, being 
equal to zero for $q\leq 1$.

\section{Conclusion}
We have provided the exact solution of a simple model of a vesicle attached to a surface subject to a pulling force
on the vesicle. 
We have used two methods of solution which illuminate the generalisation of the well-known ``kernel'' method for the 
solution of linear functional equations using a ``catalytic'' variable and result in a non-trivial identity.
The solution is expressed in terms of well-studied $q$-series and can be analysed simply. Our model has a 
non-trivial phase transition when no osmotic pressure is applied.
 
\section*{Acknowledgements}

Financial support from the Australian Research Council via its support
for the Centre of Excellence for Mathematics and Statistics of Complex
Systems is gratefully acknowledged by the authors. A L Owczarek thanks the
School of Mathematical Sciences, Queen Mary, University of London for
hospitality.

\end{document}